\documentclass[12pt]{article}
\usepackage{amssymb,graphicx}

\makeatletter
\renewcommand\section{\@startsection {section}{1}{\z@}%
                                 {-3.5ex \@plus -1ex \@minus -.2ex}
                                   {2.3ex \@plus.2ex}%
                                   {\normalfont\large\bfseries}}
\renewcommand\subsection{\@startsection{subsection}{2}{\z@}%
                                   {-3.25ex\@plus -1ex \@minus -.2ex}%
                                     {1.5ex \@plus .2ex}%
                                     {\normalfont\bfseries}}
\renewcommand\subsubsection{\@startsection{subsubsection}{3}{\z@}%
                                   {-3.25ex\@plus -1ex \@minus -.2ex}%
                                     {1.5ex \@plus .2ex}%
                                     {\normalfont\itshape}}
\makeatother

\newcommand{\Letter}{
\setlength{\textwidth}{16.5cm}
   \setlength{\textheight}{22.6cm}
    \hoffset=-0.6in
\voffset=-2.1cm }

\Letter

\newcommand{\gsim}{ \lower .75ex \hbox{$\sim$} \llap{\raise .27ex \hbox{$>$}} }
\newcommand{\lsim}{ \lower .75ex \hbox{$\sim$} \llap{\raise .27ex \hbox{$<$}} }

\begin{document}
\thispagestyle{empty}
\begin{flushright}
\parbox[t]{1.3in}{
MAD-TH-06-4\\
hep-th/0605189}
\end{flushright}

\vspace*{0.5in}

\begin{center}
{\large \bf DBI Inflation in the Tip Region of a Warped Throat}

\vspace*{0.5in} {Steven Kecskemeti, John Maiden, Gary Shiu, Bret
Underwood}
\\[.3in]
{\em
     Department of Physics,
     University of Wisconsin,
     Madison, WI 53706, USA $^{\dagger}$
         }\\[0.5in]
\end{center}

\begin{center}
{\bf
Abstract}
\end{center}

\noindent Previous work on DBI inflation, which achieves inflation
through the motion of a $D3$ brane as it moves through a warped
throat compactification, has focused on the region far from the tip
of the throat. Since reheating and other observable effects
typically occur near the tip, a more detailed study of this region
is required. To investigate these effects we consider a generalized
warp throat where the warp factor becomes nearly constant near the tip.
We find that it is possible to obtain 60 or more
e-folds in the constant region, however large non-gaussianities are
typically produced due to the small sound speed of fluctuations. For
a particular well-studied throat, the Klebanov-Strassler solution,
we find that inflation near the tip may be generic 
and it is difficult to 
satisfy
current bounds on non-gaussianity, but other throat solutions may
evade these difficulties.

\vfill
\hrulefill\hspace*{4in}

{\footnotesize $^{\dagger}$ Email:
\parbox[t]{7in}{kecskemeti@wisc.edu, jwmaiden@wisc.edu, shiu@physics.wisc.edu, bjunderwood@wisc.edu}}

\newpage

\section{Introduction}
\label{sec:introduction}

The successes of the inflationary paradigm have motivated the
construction of a number of inflationary scenarios from string
theory \cite{DvaliTye, DD, KKLMMT, AngleInflation, StringInflation,
Giant}. A recurrent set of tools in many of these constructions
involves one way or the other the idea of brane inflation
\cite{DvaliTye} and/or warped throats. In particular, significant
warping has shown to play an interesting role in getting the right
scale of inflation \cite{KKLMMT}, in achieving efficient reheating
\cite{Reheating, Kofman:2005yz}, and in the presence of multiple
throats in ensuring the stability of cosmic strings
\cite{CosmicStrings} formed at the end of brane inflation. Perhaps
the most interesting application of such warped throats is in DBI
inflation where the warping imposes a position dependent local
``speed limit" on how fast the inflaton can roll\footnote{Branes
that nearly saturate this speed limit will be called
``relativisitic".}, thus allowing inflation even for steep
potentials \cite{Silverstein:2003hf, Alishahiha:2004eh, IRDBI,
ChenNonGauss}.

Indeed, warped throats often appear in string theory in the context
of flux compactifications
\cite{Herman,Gukov:1999ya,Dasgupta:1999ss,Greene:2000gh,GKP,KKLT}.
In addition to stabilizing moduli, background fluxes back-react on
the metric and so strongly warped regions can be formed if the
fluxes have support on cycles that are localized in the compactified
space. A particularly well-studied example is the warped deformed
conifold solution of \cite{Klebanov:2000hb,Klebanov:2000nc}. The
throats are generated by turning on background fluxes along the
cycles of a conifold of a Calabi-Yau, which also smooth out the
conifold singularity into a smooth $S^3$ ``cap" at the tip where the
warp factor is approximately constant \cite{Klebanov:2000hb,
Klebanov:2000nc, Conifolds, Herzog:2001}. Far from the capped tip
the throat looks like $AdS_5\times X_5$ and most studies of DBI
\cite{Silverstein:2003hf, Alishahiha:2004eh, IRDBI, ChenNonGauss,
Shandera:2006ax} have only considered brane inflation in this
region, and assumed the constant region to be negligible.
Since reheating typically occurs when the brane reaches the tip of
the throat \cite{Reheating, Kofman:2005yz}, it is possible that the
last 60 e-folds relevant for observations may arise from inflation
in the capped region of the throat. Therefore we analyze the nearly
constant region of the throat for a generalized warped throat DBI
model, where the constant region is constructed to be large enough
that the brane will spend a significant amount of time (in terms of
e-folds) in that region during inflation. Although the existence of
inflation in a constant region is initially an assumption, we find that
inflation in the Klebanov-Strassler \cite{Klebanov:2000hb} throat seems
to satisfy this assumption for weakly warped throats\footnote{which is typically considered in the literature in order for reheating to be efficient \cite{Reheating,Kofman:2005yz}.}
and it may be realized in other throat geometries.
However, this class
of models tends to suffer from large non-gaussianities, where the
exact details depends on the construction of the throat. This does not
rule them out as viable options for inflationary scenarios, but puts
constraints on their construction in order to avoid bounds on 
non-gaussianities.
In addition, our analysis includes
how details of the geometry of the throat are encoded in the
observables such as density perturbations and non-gaussianities.
In general this information cannot be
separated from other typical slow-roll parameters (i.e. the shape of
the potential and the Hubble scale during inflation), though
specific warped throat compactifications might evade this issue.

Since not many examples of warped throats whose explicit metrics are
known, we will consider a reasonably generic form for the warp
factor describing the throat which reduces to $AdS_5\times X_5$ or
the Klebanov-Strassler solution in known
limits.  In addition to investigating the dynamics and calculating
the inflationary observables for a general warp factor we consider
two specific models of DBI inflation in the Klebanov-Strassler
warped throat, labeled by the direction of motion of the D-brane in
the throat.  In UV DBI inflation \cite{Silverstein:2003hf,
Alishahiha:2004eh} the D-brane falls into the throat towards the
tip, while in the IR DBI model \cite{IRDBI} the D-brane starts deep
in the throat near the tip and moves towards the unwarped bulk
region.

We find that a sufficient number of e-foldings can occur in the
nearly constant region near the tip in the UV model, provided one
can satisfy certain constraints regarding the development of the
tachyon or other stringy effects at the tip. A generic feature of
inflation in the constant region is that there are large non-gaussian fluctuations
produced during inflation due to the small sound speed of the
inflaton near the tip.  For the KS solution, inflation naturally
occurs in the constant region of the throat for weakly warped throats ($h_{tip}\sim 10^{-2}$)
and observational constraints
on density perturbations and non-gaussianities cannot be simultaneously satisfied.
It could be possible that for alternative warped
throat compactifications, the predictions may be consistent with
current bounds on non-gaussianities from the WMAP three year data
\cite{WMAP}.

We also study the IR scenario in the tip and AdS regions and again
find that large non-gaussianities are almost always produced for
inflation near the tip. The exception is when the D-brane
is between a critical value for the field (to be defined)
and the end of the throat:
here the motion of the inflaton is non-relativistic and
non-gaussianities are suppressed (see Section \ref{subsec:irdbi} for
more details). For non-gaussianities in IR DBI to be consistent
with observations, the last 50-60 e-folds must occur in the AdS
region of the throat and the Hubble scale during inflation must be
lower than $H\sim 10^{10}$ GeV.

The paper is organized as follows.  In Section 2 we review the DBI
inflationary scenario and the Hamilton-Jacobi approach we will be
using throughout the paper.  In Section 3 we investigate DBI
inflation near the tip of a generic warped throat and calculate the
observables. In Section 4 we review the warped throat solution of
\cite{Klebanov:2000hb, Klebanov:2000nc} and discuss an approximation to the
Klebanov-Strassler throat called the ``mass gap."  Here we also
examine the UV and IR DBI inflation models for the mass gap
approximation of the KS throat and discuss the implications. We
conclude in Section 5.  Some details about the KS throat and
non-gaussianities for a generic throat are found in Appendices A and
B, respectively.

\section{Overview of DBI Inflation}
\setcounter{equation}{0}

We will take the metric for our throat region to be of the form
\begin{eqnarray}
ds_{10}^2 = \tilde{f}(r)^{-1/2} ds_4^2 + \tilde{f}(r)^{1/2} ds_6^2
\end{eqnarray}
where $r$ is the transverse radial coordinate between the branes. As
was shown in \cite{Silverstein:2003hf}, the acceleration for
speed-limited motion we will consider is small so we can treat the
Dirac-Born-Infeld action as a good approximation to the motion of
the $D3$ brane. Rescaling the radial coordinate as $r =
\frac{\phi}{\sqrt{T_3}}$, the DBI action for the motion of the $D3$
brane as it moves through the warped throat is
\begin{eqnarray}
\label{eqn:DBIaction}
    S=-\int
    d^4x\sqrt{-g}\left(f(\phi)^{-1}\sqrt{1+f(\phi)g^{\mu\nu}\partial_{\mu}\phi\partial_{\nu}\phi}
    -V(\phi) - f(\phi)^{-1}\right).
\end{eqnarray}
Note that we have assumed that only the RR 4-form flux $C_4$ has components
along the brane and have ignored the flux due to the
NSNS 2-form $B_2$.  This is consistent
with the SUGRA solutions of \cite{Klebanov:2000hb}, where we ignore the
$B_2$ contribution because its pullback only depends on the angular
coordinates of the $X_5$ base of the throat, and we will
only be interested in the motion of the $D3$ brane along the radial
coordinate\footnote{The case of considering the angular coordinates at
the tip in a slow roll context was considered by \cite{Giant}.  It
would be interesting to see if DBI changes this scenario, and we
leave this to future work.}.

We will consider the following general form for $f(\phi)$:
\begin{equation}
f(\phi)^{-1} = f_{0}+f_{2} \phi^2 + f_{4} \phi^4.
\label{eq:GeneralF}
\end{equation}
Our choice of $f(\phi)$ is motivated by the geometry of the
Klebanov-Strassler (KS) warped throat, though we shall use this
general form of $f(\phi)$ for most of the analysis. In the limit
$f_0\rightarrow 0$ we no longer have a cutoff throat, since the warp
factor $f(\phi)\propto \tilde{f}(r)$ does not approach a constant at
the tip of the throat.  In later analysis (Section 4) we
will compare the ``AdS'' solution ($f_0 \rightarrow 0, f_2
\rightarrow 0, f_4 = \frac{1}{\lambda}$) to a ``mass gap" solution
which models the tip geometry with coefficients
\begin{equation}
f_0 = \frac{\mu^4}{\lambda},\hspace{.5in} f_2 = \frac{2
\mu^2}{\lambda} \hspace{.5in} f_4 = \frac{1}{\lambda}
\label{eq:GeneralFMassGap}
\end{equation}

Now that we have defined the general form of our metric, we consider
only spatially flat cosmologies and fields in our action,
\begin{eqnarray}
    ds_4^2&=&-dt^2+a(t)^2dx^2 \nonumber \\
    \phi &=& \phi(t)
\end{eqnarray}
and we will study the resulting FRW cosmology of the warped throat.
The Friedmann equations take the standard form
\begin{eqnarray}
\label{eqn:AdSRho}
    3H^2&=&\frac{1}{M_{p}^2}\rho \\
    2\frac{\ddot{a}}{a}+H^2&=&-\frac{1}{M_{p}^2}p
\end{eqnarray}
where $H=\frac{\dot{a}}{a}$ is the Hubble parameter (dots denote
derivatives with respect to comoving time $t$), and the energy
density $\rho$ and pressure $p$ for the DBI Lagrangian are given by
\begin{eqnarray}
\label{eqn:density}
    \rho=\frac{\gamma}{f}+(V-f^{-1})
\end{eqnarray}
\begin{eqnarray}
    p=-\frac{1}{f\gamma}-(V-f^{-1})
\end{eqnarray}
and $\gamma$ is defined as
\begin{eqnarray}
    \gamma\equiv\frac{1}{\sqrt{1-f(\phi)\dot{\phi}^2}}.
\end{eqnarray}
The Greek letter $\gamma$ was purposely used as this factor is
analogous to the Lorentz factor of special relativity. Notice that
the motion of the branes will be constrained by the position
dependent speed limit
\begin{eqnarray}
\dot{\phi}^2 \leq \frac{1}{f(\phi)}.
\end{eqnarray}
This will be important when comparing the behavior of $\phi(t)$ in
the AdS and cutoff throat geometries. Also notice that $\rho$ and
$p$ reduce to the usual expressions in the limit of small $\dot{\phi}$.

Finally, varying the DBI action with respect to the field results in
the equation of motion for $\phi$,
\begin{eqnarray}
    \ddot{\phi}+\frac{3f'}{2f}\dot{\phi}^2-\frac{f'}{f^2}+\frac{3H}{\gamma^2}\dot{\phi}+
    \left(V'+\frac{f'}{f^2}\right)\frac{1}{\gamma^3}=0,
\end{eqnarray}
where from now on a prime denotes derivative with respect to $\phi$.

\subsection{Hamilton-Jacobi Approach}
To simplify our work we shall study the action and the resulting
cosmology using the Hamilton-Jacobi formalism \cite{HamJac}. In the
Hamilton-Jacobi approach, the scalar field $\phi$ is viewed as the
time variable, thus $\phi=\phi(t)$ must be monotonic. All of our
fields ($H$, $\gamma$, $f$, $V$) from now on will be functions of
$\phi$ unless stated otherwise. Taking the time derivative of Eq.
(\ref{eqn:AdSRho}), and using the equation of motion for $\phi$, we
obtain
\begin{eqnarray}
    6HH'\dot{\phi}=-\frac{1}{M_p^2}3H\gamma\dot{\phi}^2,
\end{eqnarray}
which, after dividing both sides by $\dot{\phi}$ (permitted by the
monotonic behavior of $\phi$), results in
\begin{eqnarray}
    \dot{\phi}=-2M_{p}^2\frac{H'}{\gamma}.
\end{eqnarray}
Using the definition of $\gamma$, and solving for $\dot{\phi}$ we
have
\begin{eqnarray}\label{phidot}
    \dot{\phi}=\frac{-2H'}{\sqrt{\frac{1}{M_{p}^4}+4fH'^2}}.
\end{eqnarray}
Substituting this result back into the Friedmann equation, and using
the definition of $\rho$, a consistency condition for the potential
may be obtained,
\begin{eqnarray}\label{pot}
    V (\phi) =3 M_p^2 H^2-\frac{M_p^2}{f}\sqrt{\frac{1}{M_p^4}+4 f H'^2}+\frac{1}{f}.
\end{eqnarray}
Given a potential $V(\phi)$, we can then solve for $H(\phi)$, or
similarly, choosing an $H(\phi)$, we can find a potential that
satisfies the equations above. The latter approach is useful because once
the form of $H(\phi)$ is known, we can work backwards to calculate
$\dot{\phi}$, integrate to find $\phi$, and then use
$H=\frac{da/dt}{a}$ to integrate and find the form of the scale
factor. The disadvantage of this approach is that one must make an
ansatz for $H(\phi)$; it is often difficult to choose a functional
form that generates the desired form of the potential, however we
will see that simple choices can be made for potentials of interest
for inflation.

\section{DBI Inflation in a Generic Warped Throat}
\setcounter{equation}{0}
\label{sec:genwarpthroat}

We would like to solve Eqs.(\ref{phidot}) and (\ref{pot}) for a
string theory motivated potential, which will be a function of the
field $\phi$. Following \cite{KKLMMT} we will
take the potential to be,
\begin{equation}
V= V_0 + V_2 \phi^2 +V_4 \phi^4 - \frac{V_c}{\phi^{4}}. \label{eq:Potential}
\end{equation}
This form includes the leading renormalizable terms that are
symmetric under the $\textbf{Z}_{2}$ symmetry of the warped throat
as well as a Coulomb term which describes the attraction of a $D3$
and an $\overline{D3}$ brane. We will treat the $V_i$ terms as
(almost) arbitrary constants since their precise form will be fixed
by the details of moduli stabilization, $\alpha'$ effects, and
non-perturbative contributions to the superpotential;
$V_c$ is given
by the perturbation to the warped background from the $D3$, 
which for an asympototically
$AdS_5\times X_5$ throat is given by \cite{KKLMMT, Shandera:2006ax}
$V_c = v \frac{(T_3 h_{tip}^4)^2}{\phi^4}$, where $v$ is a geometric factor of the
$X_5$ space. Much progress has been made recently in
understanding the generation of these potentials in specific
geometries which preserves more supersymmetry (such as $K3 \times
T^2/\textbf{Z}_{2}$ and $T^{6}/\textbf{Z}_{2}$)
\cite{Shandera:2004zy}. The form of the potential is not known in
general, however, and we will sidestep the subtleties involved in
these constructions.

Previous work on the UV DBI model \cite{Silverstein:2003hf,
Alishahiha:2004eh, Shandera:2006ax} have continued the AdS region of
the throat all the way to the tip at $\phi = 0$.
Our
objective is to analyze the dynamics of the inflaton near the tip in
a more generic warped background, where the space near the tip is no
longer approximately AdS. With the introduction of the generic warp
factor in Eq. (\ref{eq:GeneralF}) our warped background now
approaches a constant ($f_0$) near the tip, which gives us a region
of nearly constant warping. Before we continue we need to make a few
assumptions about this region.

Since our analysis focuses on the region near the tip ($\phi \approx
0$), we would like to ensure that the branes are separated by more
than a (local) string length $r\geq \ell_{s} h_{tip}^{-1}$ in the
region of interest so we can ignore the development of the tachyon.
The nearly constant region of the tip is defined to be the
region where the $f_0$ term dominates the warp factor in
Eq.(\ref{eq:GeneralF}), $\phi<\phi_{tip} = \sqrt{\frac{f_0}{f_2}}$.
In order to describe inflation in the nearly constant region
using a supergravity approximation, we
require $\phi_{tip} \geq \phi_s$ where $\phi_s$ is the
value of the field corresponding to a warped string length. Using
the normalization of the inflaton field $\phi =
r \sqrt{T_{D3}}$, this translates to requiring
\begin{equation}
f_0 > \frac{m_s^2 h_{tip}^{-2}}{g_s} f_2;
\label{eq:TachyonFineTuning}
\end{equation}
since a priori there is no relation between $f_0$ and $f_2$ it is
not clear whether this is a generic feature of warped throats.
We will see later that experimental constraints on the
parameters of the mass gap solution to the Klebanov-Strassler throat
Eq.(\ref{eq:GeneralFMassGap}) satisfy this constraint 
for weakly warped throats ($h_{tip}\sim 10^{-2}$), i.e.
the inflaton will always spend a measurable amount of time (in terms
of e-folds) in the constant region of the throat.

One particular concern is that, as in slow roll inflationary models
constructed from $D\bar{D}$ pairs, for small $\phi$ the Coulomb term
in Eq.(\ref{eq:Potential}) can spoil inflation by leading to rapid
change in $\phi$. This can happen when the Coulomb term in Eq.(\ref{eq:Potential})
dominates over the mass term.  To prevent this, we will assume that the mass
term is sufficiently large that the Coulomb term doesn't dominate until
stringy effects take over,
\begin{equation}
m_{\phi}\geq \frac{\sqrt{V_c}}{\phi_s^3} = v^{1/2} h_{tip}^{7}g_s^{1/2} m_s
\label{eq:CoulombFineTuning}
\end{equation}
where in the last equality we have used the estimate for $V_c$ for
an $AdS_5\times X_5$ background from above, and inserted the field
value at which stringy effects become important.  Notice that even
for weakly warped throats ($h_{tip}\sim 10^{-2}$) this lower bound
on the mass is very weak ($\frac{m_{\phi}}{m_{s}}\geq 10^{-14}$),
and we will assume that it can be easily satisfied.  Indeed, for
masses near this lower bound we expect slow-roll inflation
\cite{Shandera:2006ax}; since our intent is to study DBI inflation
we will not be concerned with this case.  We have performed a
numerical simulation of the effects of the Coulomb term during
inflation and have found that for inflaton masses larger than the
above bound the Coulomb term is indeed negligible throughout the
region where the supergravity analysis is justified.  If the Coulomb
term does dominate the potential, inflation in the constant region
quickly ends.  However, for inflaton masses smaller than
Eq.(\ref{eq:CoulombFineTuning}) we do not expect DBI inflation at
all, so we will not consider this case in this work.

With a large mass term as expected in DBI inflation, however, one may be
worried
that the potential could violate our effective field theory description.  
In
particular,
in order to ignore stringy effects we require $V \leq  f(\phi)^{-1} m_s^4$
throughout the throat.
At the gluing the warp factor is one (the corresponding value of the field
for an asymptotically AdS
throat there is $\phi_{glue} = R_{+}/\ell_s m_s$), and after some algebra we
have the restriction
$m_s/M_p \geq \frac{R_+}{\ell_s}\frac{m_{\phi}}{M_p} \sim 10^{-2}$, where we
used $m_{\phi}/M_p \sim 10^{-5}$
as is required for sufficient inflation with the correct level of density
perturbations (shown below).
This can be challenging from the point of view of embedding the throat in a
compact space because of the large volume of the throat.
However, it is conceivable that explicit warped throats satisfying 
these constraints can be constructed (e.g., by exploring warped throats 
with small angular volume). We leave this to future work.

With these restrictions on the phase space in mind, we can now solve
Eqs.(\ref{phidot}) and (\ref{pot}) analytically.  The
Hamilton-Jacobi equations are most simply solved by choosing a
particular form for $H(\phi)$, and then using this form to solve for
the dynamics of $\phi$.  While this choice is motivated by the form
of the potential that is generated, it is also well suited to our
analysis since we are looking at late-time behavior (small $\phi$),
so we will only be interested in the leading behavior of $H$ with
$\phi$.
The choice
$H(\phi) = h_1 \phi$, is compatible with a potential of the form
Eq.(\ref{eq:Potential}) 
where the mass term dominates over all other terms, which
is what we expect for DBI inflation;
additional
powers of $\phi$ can be included (corresponding to higher order in $\phi$ terms in
\ref{eq:Potential})) but since we are interested in the
leading behavior at late times we will ignore them. A
numerical calculation of Eq. (\ref{pot}) with the full form of the
potential in Eq. (\ref{eq:Potential}) requires detailed
computational analysis that is beyond the scope of this paper, and
would be an interesting topic for future research.

To summarize, to obtain our results we are working under
four main assumptions:
\begin{itemize}
\item Near the tip of the throat, the warp
factor takes the form of Eq.~(\ref{eq:GeneralF}).
\item This constant region is larger than a warped string length away from the tip, i.e. we can ignore stringy effects
and treat this with a supergravity approximation.
\item The Coulomb term in the potential, Eq. (\ref{eq:Potential}), is
subdominant when compared to the mass term, and thus can be ignored.
\item Our choice of $H(\phi) = h_1 \phi$ is sufficient to describe the
evolution of $\phi$ near the tip. Higher order terms can be dropped
since we looking at the region where $\phi \rightarrow 0$.
\end{itemize}
These assumptions are motivated and, as we will see, are satisfied by the
Klebanov-Strassler throat.

Working with these assumptions, the consistency condition
Eq. (\ref{pot}), together with Eq. (\ref{eq:GeneralF}) generates a
potential with the following coefficients
\begin{eqnarray}
    V_0&=& f_0 (1- \frac{2 h_1 M_{p}^2}{A})
    \\ \nonumber
    V_2&=& f_2 (1- \frac{2 h_1 M_{p}^2}{A}) + 3 h_1^2 M_{p}^2 + \frac{f_2 h_1 M_{p}^2 A}{f_0}
    \\ \nonumber
    V_4&=& f_4(1-\frac{2 h_1 M_{p}^2}{A}) + \frac{f_2^2 h_1 M_{p}^2 A^3}{4 f_0^3} +\frac{f_4 h_1 M_{p}^2 A}{f_0},
\end{eqnarray}
where we have defined the combination
\begin{equation}
A = \frac{f_0^{1/2}}{\sqrt{1+\frac{f_0}{4 h_1^2 M_{p}^4}}}
\end{equation}
 From our $V_2$ term, we can now solve for the constant $h_1$ to
obtain
\begin{eqnarray}
    h_1\approx \frac{m_{\phi}}{\sqrt{6} M_{p}},
\label{eq:h1}
\end{eqnarray}
where we let $V_2 = \frac{1}{2} m_{\phi}^2$ and assumed $m_{\phi}\gg
f_2$; this assumption is reasonable since the
inflaton mass must be large in order to trust our analysis near the
tip and to be able to ignore the Coulomb term. We can rewrite the
cosmological constant term $V_0$ as
\begin{eqnarray}
    V_0\approx f_0(1- \sqrt{\frac{2}{3}}\frac{m_{\phi} M_{p}}{A});
\end{eqnarray}
as mentioned above, we will consider $V_0$ to be a tunable parameter
that can be set to this value. In practice, this consistency
condition for $V_0$ is not important and is an artifact of the
Hamilton-Jacobi method; the dynamics for the inflaton
field are qualitatively similar as long as the mass term dominates the
potential.

To solve the remaining equations of motion we put our general form
for the warp factor into Eq. (\ref{phidot}), which gives us, for
small $\phi$
\begin{equation}
\label{eqn:latephi}
\dot{\phi} = -(A + \frac{1}{2}\frac{f_2 A^3}{f_0^2} \phi^2) + {\mathcal{O}}(\phi^4);
\end{equation}
this leads to a late time behavior
\begin{eqnarray}
\label{eqn:phimassgap}
\phi(t) = \sqrt{\frac{2}{f_2}}\frac{f_0}{A}
            \tan\left[\frac{\sqrt{\frac{1}{2}f_2}A^2}{f_0}(t_f-t)\right]
\end{eqnarray}
where
\begin{equation}
t_f \equiv \frac{f_0}{\sqrt{\frac{1}{2}f_2}A^2}\arctan\left[\sqrt{\frac{f_2}{2}}\frac{A}{f_0}\phi_0 \right],
\end{equation}
$\phi_0$ is the initial starting point of the brane, and $t_f$ is
defined by $\phi(t_f)=0$. Note that this form for $\phi(t)$ is still
consistent with our requirement that $\phi(t)$ be monotonic; for the
above values of $\phi(t)$ the function goes through less than half a
period.

The generic late-time behavior of this solution is different from
the previously observed behavior in an AdS background
\cite{Silverstein:2003hf} since for late times,
\begin{equation}
\phi(t) \approx A (t_f-t).
\end{equation}
Notice that the inflaton reaches the origin in a finite time, as
would be expected for a finite throat. This is to be compared with
the AdS solution in which $\phi(t) \rightarrow \sqrt{\lambda}/t$ at
late times. The AdS solution can be obtained from
Eq.(\ref{eqn:phimassgap}) through the limit $f_0, f_2\rightarrow 0$ in
Eq.(\ref{eq:GeneralF}) and choosing the appropriate late
time behavior for $t_f$.

 From our solution Eq.(\ref{eqn:phimassgap}) and using
our ansatz $H=h_{1}\phi$ a straightforward analysis gives the scale factor and number
of e-folds as,
\begin{eqnarray}
a(t) &=& a_{0} \left[\cos\left(\frac{\sqrt{\frac{1}{2}f_2} A^2 (t-t_f)}{f_0} \right)\right]^{\alpha} \nonumber \\
N_{e} &=& \alpha
\log\left[\frac{\cos\left(\frac{\sqrt{\frac{1}{2}f_2} A^2 (t_e-t_f)}{f_0}\right)}
{\cos\left(\frac{\sqrt{\frac{1}{2}f_2} A^2 (t_0-t_f)}{f_0}\right)}\right]
\label{eq:scalefactorEfolds}
\end{eqnarray}
where the exponent is $\alpha = \frac{2 h_1 f_0^2}{f_2 A^3}$; $t_0$
is the initial time and $t_e\leq t_f$ is the time that inflation ends
and is defined by $\phi(t_e) = \sqrt{\frac{A}{h_1 -
\frac{f_2 A^3}{2 f_0^2} }}$. Depending on the model, inflation may
end before the branes annihilate at the tip of the throat. Our
results can be made more transparent by noting the scale factor and
number of e-folds can be written in terms of $\phi(t)$ as follows,
\begin{eqnarray}
a(t) &=& a_0 (A+\frac{1}{2}\frac{f_2 A^3}{f_0^2}\phi(t)^2)^{-\alpha/2} \nonumber \\
N_{e} &=& \frac{\alpha}{2} \ln\left[\frac{1+\frac{1}{2}\frac{f_2
A^2}{f_0^2}\phi(t_i)^2}{1+\frac{1}{2}\frac{f_2
A^2}{f_0^2}\phi(t_e)^2}\right] \approx \frac{\alpha}{2}\log(2).
\label{eq:EFoldsGeneral}
\end{eqnarray}
We simplified the expression for the number of e-folds since we are
interested in inflation starting near the beginning of the tip
region, $\frac{f_2}{f_0}\phi_i^2 \approx 1$ (see the discussion at
the beginning of this section) and ending deep in the tip region,
$\frac{f_2}{f_0}\phi_e^2 \ll 1$.  Using the definition of $\alpha$,
we note that in order to get at least 60 e-folds,
\begin{equation}
f_2\ll \frac{m_{\phi}}{M_{p}} f_0^{1/2}.
\label{eq:EfoldsTuning}
\end{equation}
This appears to be in agreement with our earlier constraint on $f_2$
needed to trust our analysis at the tip. Whether Eq.(\ref{eq:TachyonFineTuning}) or
(\ref{eq:EfoldsTuning}) is more stringent depends on the details for the specific
model.

\subsection{Inflationary Observables}
\label{subsec:geninflobs}

Now that we have shown that we can generate at least 60 e-folds near
the tip in a generically warped throat, we need to see how the tip
dynamics affect the cosmological observables. Since we are using the
Hamilton-Jacobi formalism, it is useful to define a different set of
inflationary parameters that we will use to calculate these
observables. We define our DBI analogy to the slow roll parameters
in terms of the Hubble parameter $H(\phi)$, where we start with
$\epsilon_D$, defined as
\begin{eqnarray}
\frac{\ddot{a}}{a} = H^2(\phi) (1 - \epsilon_D)
\end{eqnarray}
For inflation to occur we must have $0 < \epsilon_D < 1$. Defining
the rest of our inflationary parameters
using the conventions of
\cite{Shandera:2006ax}
\begin{eqnarray}
\label{eq:DBIEpsilon} \epsilon_D &\equiv& \frac{2
M_p^2}{\gamma(\phi)} \left( \frac{H'(\phi)}{H(\phi)} \right)^2 \\
\label{eq:DBIEta} \eta_D &\equiv& \frac{2 M_p^2}{\gamma(\phi)}
\left( \frac{H''(\phi)}{H(\phi)} \right) \\
\label{eq:DBIKappa} \kappa_D &\equiv& \frac{2 M_p^2}{\gamma(\phi)}
\left( \frac{H'(\phi)}{H(\phi)} \frac{\gamma'(\phi)}{\gamma(\phi)}
\right)
\end{eqnarray}
These are, to leading order in $\phi$,
\begin{eqnarray}
\epsilon_D &=& \frac{2 M_p^2}{\phi^2 \gamma} \approx
\frac{1}{h_1}(\frac{A}{\phi^{2}}
       +\frac{1}{2}\frac{f_2 A^3}{f_0^2}) + {\cal{O}}(\phi^2) \\
\eta_D &=& 0  \\
\kappa_D &=& \frac{2 M_p^2}{\phi} \frac{\gamma'}{\gamma^2} \approx
-\frac{f_2 A^3}{h_1 f_0^2} +{\cal{O}}(\phi^2).
\end{eqnarray}

The scalar spectral index $n_s - 1 = \frac{d\ln {\cal{P}}_R}{d
\ln k}$ is given by,
\begin{eqnarray}
n_s - 1 &=& -4\epsilon_D + 2\eta_D -2\kappa_D \nonumber  \\
        &=& -\frac{4 A}{h_1 \phi^{2}} + {\cal{O}}(\phi^{2}) \approx -\frac{4 \log(2)}{N_e}
\end{eqnarray}
Note that this gives us a red shifted spectral index for non-vanishing
$f_0,f_2$; this is in contrast to DBI in the AdS throat where
$n_s =1$ to all orders in the inflationary parameters \cite{Shandera:2006ax}.

The tensor mode spectral density is
\begin{eqnarray}
{\cal{P}}_h = \frac{2H^2}{\pi^2 M_p^2},
\end{eqnarray}
the corresponding tensor index is
\begin{eqnarray}
n_t &\approx& -2 \epsilon_D \nonumber \\
        &=& -\frac{2A}{h_1 \phi^2}-\frac{f_2 A^3}{h_1 f_0^2} +{\cal{O}}(\phi^2) \approx -\frac{2\log(2)}{N_e}
\end{eqnarray}
and ratio of power in tensor modes to scalar
modes,
\begin{eqnarray}
r &=& \frac{16 \epsilon_D}{\gamma} \nonumber \\
&=& \frac{8}{h_1^2 M_{p}^{2}}(\frac{A^2}{\phi^{2}}+ \frac{f_2
A^4}{f_0^2}) +{\cal{O}}(\phi^2) \\ \nonumber &=& \frac{8}{h_1^2
M_{p}^{2}}(\frac{f_2 A^2}{f_0}+ \frac{f_2 A^4}{f_0^2})
\end{eqnarray}
where in the last line we have evaluated $r$ at $\phi_{tip}$ (recall
$\phi_{tip} = \sqrt{\frac{f_0}{f_2}}$ is the boundary between the constant
and non-constant regions of the throat). The
running of the spectral indices are
\begin{eqnarray}
\frac{d n_s}{d \ln k} &\approx& \frac{d}{dN_e}(4\epsilon_D - 2 \eta_D + 2 \kappa_D)  \nonumber \\
&=& -\frac{8 A^2}{h_1^2 \phi^4}-\frac{4 f_2 A^4}{h_1^2 f_0^2
\phi^2}- \frac{2A^4 (-4 f_0 f_2^2 + 4 f_0^2 f_4 + 3f_2^2 A^2)}{f_0^4 h_1^2} +{\cal{O}}(\phi^2) \nonumber \\
&\approx& -\frac{8\log(2)}{N_e^2} \\
\frac{d n_t}{d\ln k} &\approx & 2\frac{d ~\epsilon_D}{dN_e} \nonumber \\
&=& -\frac{4 A^2}{h_1^2 \phi^4}-\frac{2 f_2 A^4}{h_1^2 f_0^2 \phi^2} \nonumber \\
&\approx& -\frac{4\log(2)}{N_e^2}\\
\end{eqnarray}
where we used $\frac{d}{dN_e} = -\frac{\dot{\phi}}{H}\frac{d}{d\phi}
= \frac{2 M_{p}^2 H'}{\gamma H}\frac{d}{d\phi}$ for derivatives with
respect to the number of e-folds.

The level of non-gaussianities up to leading powers of $\gamma$
\cite{Chen:2006nt} is given by\footnote{This result is generic to
DBI inflation and independent of the choice of $f$ and $H$. See
Appendix \ref{sec:nongauss} for more details.},
\begin{eqnarray}
\label{eq:NonGauss} f_{NL} &\approx& 0.32 \gamma^2 \\ \nonumber
&\approx& 0.32 \left(1 + \frac{4h_1^2 M_p^4}{f_0} \right),
\end{eqnarray}
and the level of density perturbations is given by,
\begin{eqnarray}
\delta_{H} &=& \frac{\delta \rho}{\rho} = \frac{H}{M_{p}}\frac{1}{\sqrt{\epsilon_D c_s}} \nonumber \\
  &=& \frac{\sqrt{2} h_1^2 \phi^2}{A}
  \\ \nonumber &=& \frac{\sqrt{2} h_1^2 f_0}{f_2 A} = \frac{\sqrt{2}}{\log 2}\frac{f_2}{f_0^{1/2}}N_e^2
\end{eqnarray}
We see that in order to get the right level of density perturbations
we must choose
\begin{equation}
f_2\approx \frac{\delta_{H}}{N_e^2}f_0^{1/2}
\label{eq:DensityFineTuning}
\end{equation}
Comparing this to Eq.(\ref{eq:EfoldsTuning}), once we have set the
density perturbations at the right level, in order to get enough
e-folds near the tip we have the requirement $\frac{m_{\phi}}{M_p}
\gg \frac{\delta_H}{N_e^2}$. Since the right hand side can be quite
small, this is not very stringent tuning on the inflaton mass.
Indeed, the implicit lower bound on $m_{\phi}$ from the Coulomb term
in Eq.~(\ref{eq:CoulombFineTuning}) may be more restrictive in
general.

Because the dynamics of the inflaton are different near the tip of a
cutoff throat in comparison to the pure AdS throat, we would like to be able to
measure in some way the warp factor at the tip, $f_0$; we see that the
combination,
\begin{equation}
\frac{f_0}{M_{p}^4} \approx \frac{r^2 \delta_{H}}{128}
\end{equation}
will allow us to make such a measurement.

However, the primary concern for DBI inflation models with a cutoff
throat is that the non-gaussianities, Eq.(\ref{eq:NonGauss}), are
generically too large as a result of their small sound
speed (In DBI inflation the sound speed $c_s =
\frac{1}{\gamma}$). In particular, the current bound on
non-gaussianities from the WMAP three year data set \cite{WMAP}
constrains $-54 < f_{NL} < 114$. For throats of the Klebanov-Strassler type,
$f_0 = h_{tip}^4 m_s^4$, so we can write
\begin{equation}
f_{NL}\approx
0.32\left(\frac{M_p}{m_s}\right)^4\left(\frac{m_{\phi}}{M_p}\right)^2
h_{tip}^{-4} \approx \left(\frac{10^{-6}}{G \mu_{cs}} \right)^2~,
\label{eq:NonGaussThroat}
\end{equation}
where we used $\mu_{cs} \approx m_{s}^2 h_{tip}^2$ as the tension of cosmic
strings produced at the tip.  This implies
that to keep cosmic strings consistent with observational bounds ($G\mu_{cs}
\leq 10^{-6}$) the non-gaussianities may be observable, $f_{NL}\sim
{\mathcal O}(1)$.
With $h_{tip}\sim 10^{-2}$ and $m_s\sim (10^{-2}-10^{-3})M_{p}$, even for inflaton masses much smaller
than the Planck scale the non-gaussianities will be quite large.  However, an $f_0$ with
a different dependence on $h_{tip}$ could potentially have a level of non-gaussianity
consistent with observations.

\section{Warped Compactifications}
\label{sec:warpcomp}
\setcounter{equation}{0}

We will now consider the specific case of the Klebanov-Strassler
(KS) throat\cite{Klebanov:2000hb,Klebanov:2000nc}.\footnote{Compact
models containing such throats have been discussed in \cite{GKP},
and the corresponding effective field theory has been explored in
\cite{WarpedEFT}.} Our setup is a Type IIB flux compactification on
a Calabi-Yau (CY) 3-fold with NS-NS and R-R fluxes turned on along
the internal compact dimensions. As in
\cite{Klebanov:2000hb,Klebanov:2000nc}, by turning on fluxes on the
cycles associated with a conifold one can stabilize the dilaton and
all the complex structure moduli.  The fluxes generate a strongly
warped ``throat'' due to their induced $D3$ charge which is glued to
the bulk CY compact space.  The fluxes are quantized by:
\begin{eqnarray}
\frac{1}{2\pi\alpha'}\int_{A}F_{(3)} &=& 2\pi M \nonumber \\
\frac{1}{2\pi\alpha'}\int_{B} H_{(3)} &=& -2\pi K,
\label{eq:FluxQuant}
\end{eqnarray}
where $A$ and $B$ are the cycles on which the fluxes are supported.
The throat is a warped deformed conifold where the deformation
replaces the conifold singularity with an $S^{3}$ ``cap".  The
metric of the warped deformed conifold is \cite{Conifolds,Herzog:2001} (notice
that our notation differs slightly from the literature):
\begin{eqnarray}
ds_{10}^2 = \tilde{f}^{-1/2} (\tau) \eta_{\mu \nu} dx^{\mu} dx^{\nu} +
\tilde{f}^{1/2} (\tau) ds_6^2
\end{eqnarray}
where $\tau$ is a coordinate along the throat
and the warp factor $\tilde{f}(\tau)$ is defined by
\begin{eqnarray}
\tilde{f}(\tau) &=& 2^{2/3} (g_s M \alpha')^2 \epsilon^{-8/3} I (\tau) \\
I(\tau) &=& \int_{\tau}^{\infty} \frac{x\coth(x)-1}{\sinh^2(x)}
(\sinh(2x)-2x)^{1/3}.
\end{eqnarray}
The parameter $\epsilon^{-2/3}$ has units of energy and describes
the deformation of the conifold, which is determined by
$\sum_{i=1}^{4} w_{i}^2 = \epsilon^2$, where $w_i$ describe the
complex structure. The undeformed conifold appears in the limit
$\epsilon\rightarrow 0$. Near the tip of the throat ($\tau =0$),
which is the region we will be interested in,
\begin{equation}
I(\tau \rightarrow 0) = a_0 + a_1 \tau^2
\end{equation}
where $a_0$, $a_1$, $\sim {\cal{O}}(1)$. We see that the warp factor
approaches a constant near the tip,
\begin{equation}
\tilde{f}(\tau) = (g_s M_p \alpha')^2 \epsilon^{-8/3} a_0 +
{\cal{O}}(\tau^2) = e^{-8\pi K/3 M g_s}+ {\cal{O}}(\tau^2).
\end{equation}

Far from the tip of the throat the geometry looks like an AdS$_5\times S^5$
throat with an exact conifold,
\begin{equation}
ds^{2}_{10}=\tilde{f}(r)^{-1/2}(\eta_{\mu\nu}dx^{\mu} dx^{\nu})
+\tilde{f}(r)^{1/2}(dr^{2}+r^{2}ds^{2}_{X_{5}}),
\end{equation}
and the warp factor takes its AdS form,
\begin{equation}
\tilde{f}(r) = \frac{\tilde{\lambda}}{r^4},
\label{eq:AdSWarpFactor}
\end{equation}
where $\tilde{\lambda} = R_+^4 =
\frac{27\pi}{4}~\ell_{s}^{4}g_{s}(MK)$, $R_+$ is the AdS length
scale, and $MK = N$ is the induced $D3$-brane charge number from the
fluxes. Previous studies of DBI inflation
\cite{Silverstein:2003hf,Alishahiha:2004eh,Shandera:2006ax} have
considered the motion of a D-brane in an AdS$_5 \times S^5$ throat.
While this metric well describes the geometry of a throat generated
by a stack of $D3$-branes and provides a good approximation to the KS
throat if one assumes the significant dynamics of the system occur
far from the tip, it does not describe the KS throat near the tip
where the warp factor becomes nearly constant.  In particular,
previous studies make use of asymptotic behavior of the inflaton in
the near horizon limit, and it is not clear that these assumptions
are applicable to the case of the warped deformed conifold.

\begin{figure}[t]
\begin{center}
\includegraphics[scale=.6]{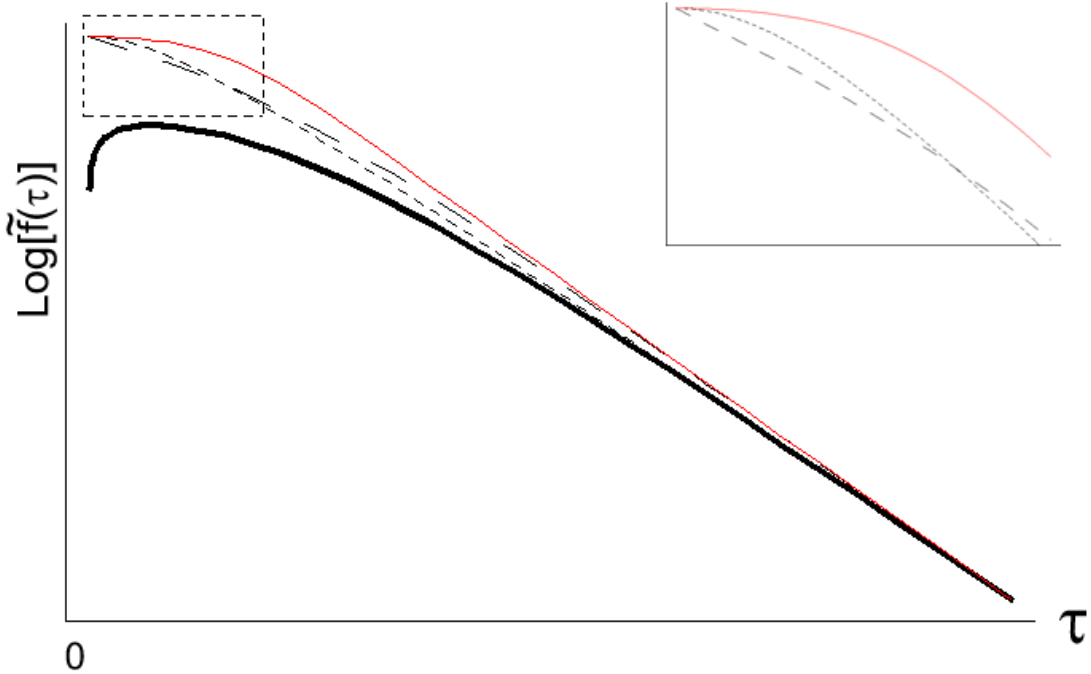}
\caption{\small Plotted are the warp factors for different throat
geometries as a function of $\tau$: the long dashed line is an
AdS$_5$ geometry, the thick line is the warp factor for a KS throat
with a log-correction as in \cite{Kofman:2005yz,KKLMMT,Herzog:2001,Klebanov:2000nc},
the thin red line
is the mass gap approximation, and the short dashed line is the
exact KS warp factor. {\it Inset}:\  The region near the tip, $\tau\sim 0$
is enlarged to show the differences in the warp factors.  Notice that the
mass gap approximation models the flattening of the KS warp factor.} \label{fig:WarpCompare}
\end{center}
\end{figure}

One possible way to remedy this, suggested
by \cite{Silverstein:2003hf}, is to use a ``mass gap" form for the
warp factor,
\begin{equation}
\tilde{f}(r) = \frac{\tilde{\lambda}}{(r^2+\tilde{\mu}^2)^2}.
\label{eq:MassGapWarp}
\end{equation}
Notice that in this model, the warp factor is approximately AdS in
the region $r \gg\tilde{\mu}$ (close to the gluing to the
Calabi-Yau). Towards the tip, as $r$ becomes very close to zero, the
warp factor is nearly constant, $\tilde{f}^{-1}\approx
\tilde{\mu}^{4}/\tilde{\lambda}$. The mass gap parameter
$\tilde{\mu}$ is chosen to give the correct warp factor at the tip
($\phi=0$) such that $\tilde{\mu} = R_+ h_{tip}$, where $h_{tip} =
e^{-2\pi K/3 M g_s}$.  We have provided a comparison of the AdS warp
factor, the mass gap warp factor, the KS warp factor, and a
log-corrected warp factor $\tilde{f}(r) =\frac{1}{r^4}(R_+^4 + 4
R_{-}^4 \log(\frac{r}{R_+}))$ as in
\cite{Kofman:2005yz,KKLMMT,Herzog:2001,Klebanov:2000nc} in Figure
\ref{fig:WarpCompare}, where we have used the relation $r-r_{tip} =
\frac{\epsilon^{2/3}}{6^{1/2}}\int_0^\tau \frac{d\tau'}{K(\tau')}$
to plot all the warp factors using the $\tau$ coordinate
\cite{Kodama} (note $K(\tau)$ is defined in Eq.(\ref{eq:Ktau}).
Note the behavior of the different warp factors for
small $\tau$: both the mass gap and the KS warp
factors\footnote{Since the log-corrected warp factor diverges for
finite $\tau$ one must set the warp factor to a constant at the tip
to accurately model the KS throat.} level out to a finite value near
the tip while the AdS warp factor does not.

While the mass gap warp factor does not satisfy the supergravity
equations of motion, we will use it merely as an analytical tool to
investigate the behavior of the more complicated Klebanov-Strassler
throat.  Since they share the same qualitative features, we will
use the
simpler mass gap for many of our analytic calculations; a brief
analysis of the KS throat can be found in Appendix \ref{sec:KSthroat} where
we show
that the mass gap solution has the same qualitative behavior near the tip.

\subsection{AdS$_5$ Throat}

We will review here the results for DBI inflation in an AdS$_5$
throat \cite{Silverstein:2003hf,Alishahiha:2004eh}.  Since the mass
gap warp factor (Eq.(\ref{eq:MassGapF})) looks like AdS
(Eq.(\ref{eq:AdSWarpFactor})) for large $r$, one can consider AdS
space to be the geometry of the throat far from the tip. The warp
factor for the AdS space is \cite{RS}
\begin{equation}
f(\phi) = \frac{\lambda}{\phi^4}
\end{equation}
where $\lambda = \frac{\tilde{\lambda}}T_3 \sim N$, and $N$ is the $D3$ charge
generating the throat.

We will consider the same form of the potential as in
Eq.(\ref{eq:Potential}), and choose $H=h_1 \phi$.  Using
Eq.(\ref{phidot}) for the AdS warp factor for small $\phi$ we have
\begin{equation}
\dot{\phi}\approx -\frac{\phi^2}{\sqrt{\lambda}}
\end{equation}
which gives us a late time solution of $\phi(t)\rightarrow
\frac{\sqrt{\lambda}}{t}$. It should be noted that the general
solution, Eq.(\ref{eqn:phimassgap}), reduces to this solution in the
limit $f_0 = \mu^4/\lambda$, $f_2 = 2\mu^2/\lambda$ and
$\mu\rightarrow 0,\ t_f\rightarrow \infty$.  (The latter is required
because the AdS solution does not reach $\phi=0$ in a finite time.)
A similar calculation as done in Section \ref{subsec:geninflobs}
will yield the number of e-folds $N_e\approx h_1 \sqrt{\lambda}$,
and the level of density perturbations $\delta_H \approx
\frac{N_e^2}{5\pi \sqrt{\lambda}}$. The rest of the observables
follow a similar pattern as written above when written in terms of
$N_e$ and we will not be concerned with their details here.

In particular, one should note that in order to get the right level
of density perturbations for $N_e\sim 60$, $\lambda \sim
10^{14}$\cite{ChenNonGauss}.  When viewed as a requirement on the
number of $D3$ charges required to generate the throat this seems
quite fine tuned ($N \sim 10^{14}$), however when viewed as a
hierarchy between the radius of the $S_3$ at the tip of the throat
and the string scale, it only requires $\lambda^{1/4}\sim R_+/\ell_s
\sim 10^3-10^4$.

\subsection{Mass Gap}
\label{subsec:massgap}

As mentioned above, the mass gap form for the warp factor is
approximately AdS$_5$ at large distances and constant near the tip.
In terms of the inflaton field $\phi$ we have,
\begin{equation}
f(\phi) = \frac{\lambda}{(\phi^2+\mu^2)^2}
\label{eq:MassGapF}
\end{equation}
where $\lambda$ is the same as in the AdS case and
$\mu = \frac{R_+}{\ell_s^2 g_s^{1/2}} h_{tip}$, as can be
seen by requiring the warp factor Eq.(\ref{eq:MassGapWarp}) to be equal
to the warping at the tip, and changing variables from $r$ to $\phi$.  The
mass gap in terms of the variables $f_0$, $f_2$, $f_4$ is given earlier in
Eq.(\ref{eq:GeneralFMassGap})\footnote{Note that we can also write
$f_0$, $f_2$, and $f_4$ in terms of the original KS warp factor
parameters from Eq.(\ref{eq:KSDBIScalar}) for small $\tau$.  However, since
the mass gap is a good
approximation to the behavior of the exact warp factor we will
limit our discussion to the former.}.

Now that we have a specific model for the cutoff throat we can
evaluate the constraint Eq.(\ref{eq:TachyonFineTuning}) to evaluate
whether stringy effects are important for our analysis; in
particular we find $h_{tip}\geq
\left(\frac{\ell_s}{R_+}\right)^{1/2}$.  Requiring $h_{tip}\leq
10^{-2}$ (this is to guarantee that there is a warped region in the
compactification) we have $\frac{R_+}{\ell_s}\geq 10^4$.  Amazingly
this is not too different than the tuning of $\lambda$ needed to
obtain the correct value of density perturbations in the AdS model.
This suggests that in order to get the right level of density
perturbations for DBI inflation in the AdS region of the throat,
then $\mu$ must also be large since $\mu \sim \lambda^{1/4}$. Since
our constant region is large, the brane spends a significant amount
of time in that region and \emph{inflation also naturally happens in
the constant region of the tip}.

One can also verify that even for weakly
warped throats ($h_{tip}\sim 10^{-2}$) the lower bound on the inflaton
mass coming from the Coulomb term Eq.(\ref{eq:CoulombFineTuning}) is
$m_{\phi}/m_{s}\geq 10^{-14}$ and is easily satisfied.
For much more strongly warped
throats
($h_{tip} \ll 10^{-2}$), this requires the radius of the AdS throat to be
large (requiring $N > 10^{14}$ D-branes).

Plugging the mass gap solution into the Hamilton-Jacobi consistency
equation Eq.(\ref{pot}), we find,
\begin{eqnarray}
V_0 &=& \frac{\mu^4}{\lambda}(1-
\frac{2 M_{p}^2 h_1\sqrt{\lambda}}{\mu^{2}}\sqrt{1+\frac{\mu^{4}}{4h_{1}\lambda M_{p}^4}})
\nonumber \\
V_2 &=& 3h_1^2 M_p + \frac{2 \mu^2}{\lambda} -
\frac{2(2h_1^2 M_p^2 \lambda + \mu^4)}{\lambda \mu^2 \sqrt{1+ \frac{4h_1^2 M_p^2 \lambda}{\mu^4}} } \\
\nonumber
V_4 &=& \frac{1}{\lambda}.
\end{eqnarray}
As we will see below, inflation requires $h_{1} \sqrt{\lambda}\sim N_e$ to
be sufficiently large, which means that Eq.(\ref{eq:h1}) requires
that the inflaton mass $m_{\phi}$ not be too small. For
$\lambda \sim 10^{14}$ this requires a tuning of
$\frac{m_{\phi}}{M_p} \geq 10^{-5}$.  Notice that this is close
to the typical Hubble-scale induced mass for GUT scale inflation
$H\sim 10^{14}\ GeV$. Together with the
requirement that $\mu<M_{p}$, which is needed in order to have a
warped throat, suggests that the $V_0$ coming from the coupling of
the scalar field to gravity must be negative. This somewhat odd
result is due to the fact that for the warped throat the warp factor
$f$ approaches a constant at the tip of the throat: from Eq.
(\ref{eqn:density}), our energy density $\rho$ obtains a positive
constant contribution from the kinetic energy that must be canceled
by a negative $V_0$ in order to satisfy our ansatz $H=h_1 \phi$.
Therefore our negative $V_0$ is an artifact of our use of
the Hamilton-Jacobi formalism. As mentioned before, this does not
trouble us because
we have numerically simulated the
equations of motion for a small, positive $V_0$ term
and found no change to the DBI speed-limited behavior of the inflaton.

In terms of the mass-gap parameters our behavior for
$\phi(t)$ is
\begin{equation}
\phi(t) = \mu\sqrt{1+\frac{\mu^4}{4h_1^2 \lambda M_{p}^4}}
\tan\left[\frac{\mu (t_f-t)}{\sqrt{\lambda} (1+\frac{\mu^4}{4 h_1^2 \lambda M_{p}^4}}\right].
\end{equation}
As mentioned above, in the limit $\mu \rightarrow 0$,
$\phi\rightarrow \mu \tan(-\mu t/\sqrt{\lambda} +\pi/2)\rightarrow \sqrt{\lambda}/t$,
and we regain the
late time behavior for the AdS warp factor, where the inflaton takes an
infinite amount of time to reach the origin.

Since the AdS and mass gap throats have different behavior for $\phi(t)$ near the tip
they also have different behaviors for
the gamma factors as a function of the inflaton,
\begin{eqnarray}
\gamma (\phi) &\rightarrow& \frac{2 h_1 \sqrt{\lambda}
M_p^2}{\phi^2} \quad \quad \mbox{AdS Warp Throat} \\
\gamma(\phi) &\rightarrow& \frac{2 M_{p}^2 h_1\sqrt{\lambda}}{\mu^2
+ \phi^2} \hspace{.5in} \mbox{Cutoff Throat}
\end{eqnarray}
For the AdS solution $\gamma$ becomes infinite as $\phi \rightarrow
0$, but in the mass gap solution it is finite and large.  This will
have implications for the non-gaussianities since $f_{NL}\propto
\gamma^2$.

The number of e-folds for the mass gap solution is given by
Eq.(\ref{eq:EFoldsGeneral}) with the appropriate values for $f_0$
and $f_2$: $N_e \approx h_1\sqrt{\lambda} \frac{\log(2)}{2}$. This
is approximately the same expression for the number of e-folds in
the AdS case, so fixing $h_1 \sqrt{\lambda}\sim {\cal O}(100)$ will
give inflation in both the AdS and constant part of the throats.
Similarly, evaluating the density perturbations at $\phi\approx
\mu$, we have the same expression for the density perturbations,
$\delta_H\approx \frac{N_e^2}{\sqrt{\lambda}}$, so yet again fixing
the density perturbations for the AdS part of the throat also fixes
them for the constant region as well. To summarize, to fit
experiment results we need a large $\lambda$, which forces $\mu$ to
be large. This is significant because it means that
\emph{even if inflation begins in the AdS region (as opposed
to starting in the nearly constant region) then the last 60 e-folds
of inflation will always be produced in the nearly constant
region}.

As noted for the general warp factor analysis, however, the primary
conflict with observations comes from the non-gaussianities (Eq.
(\ref{eq:NonGauss})),
\begin{equation}
f_{NL}\approx 0.32 \frac{M_{p}^4}{\mu^4} N_{e}^2.
\end{equation}
Since we require $\mu<M_{p}$ in order to trust our supergravity
analysis it appears that large non-gaussianities are predicted for
inflation near the tip. For $N_e \sim 60$ we cannot satisfy
experimental results for density perturbations and non-gaussianities
simultaneously.

\subsection{IR DBI Inflation}
\label{subsec:irdbi}

The model discussed above is known as UV DBI inflation because the
inflaton moves through the throat from the UV end (large $r$) to the
IR end (small $r$).  This naturally happens in string constructions
because, as we have seen, throats generated by fluxes have induced
$D3$ charges at their tips; $\overline{D3}$ branes are then
naturally attracted to the tip of the throat where they
remain\footnote{The $\overline{D3}$ branes do not annihilate with
the $D3$ charge at the tip because the charges are separated by a
potential barrier\cite{FluxAnnihilation}; the tunneling time is very
small if the flux numbers are large, so the lifetime of the state
can be tuned to be arbitrarily large.}.  $D3$ branes are then
attracted to the $\overline{D3}$ at the tip of the throat, leading
to the above action and inflation.

A variation on this model is to start the $D3$ brane at the tip of the
throat \cite{IRDBI}.  This can arise when $p$ $\overline{D3}$ branes
annihilate with the $N$ flux generated $D3$ charges and produce $N-p$ $D3$
branes at the tip.  If a $\overline{D3}$ brane resides in another
throat the attractive potential between the branes will pull the
brane out of the throat.  In particular, the potential will be of
the form,
\begin{equation}
V = V_0 - \frac{1}{2} m_{\phi}^2 \phi^2 = V_0- \frac{1}{2} \beta H^2 \phi^2;
\label{eq:IRPotential}
\end{equation}
notice that the mass term is negative, giving the direction of the
force which is pulling the $D3$ out of the throat; $\beta\approx 1$
generically. The brane then moves from the IR end of the throat to
the UV end of the throat. The moduli potential $V_0$ for the branes
will be the dominant term that drives inflation. The novel part of
this scenario is that the brane begins at the tip of the throat
where the warp factor is approximately constant, so the geometry is
important for inflation\footnote{If one is able to get 60 e-folds in
subsequent AdS region of the throat, however, the tip geometry
becomes less important.}. Inflation in the IR model is obtained in a
similar way as the UV model: the speed limit from the DBI action
restricts how fast the inflaton can roll, so even for
``relativistic" motion of the inflaton the potential stays
approximately constant and inflation can proceed.

In the IR DBI model during inflation the potential dominates over
the ``kinetic terms" in Eq. (\ref{eqn:density}), which gives us the
following Hubble factor
\begin{eqnarray}
H(\phi) &=& \sqrt{\frac{V_0- \frac{1}{2} \beta H^2
\phi^2}{3M_{p}^2}}\approx
\sqrt{\frac{V_0}{3M_{p}^2}} \\
H'(\phi) &\approx& -\frac{\beta H \phi}{6 M_p^2},
\end{eqnarray}
which combined with the Hamilton-Jacobi equations of motion give,
\begin{eqnarray}
\gamma &=& \sqrt{1 + \frac{\beta^2 H^2 \phi^2}{9(f_0 + f_2 \phi^2 +
f_4 \phi^4)} } \\
\dot{\phi} &=& \sqrt{\frac{f_0 + f_2 \phi^2 + f_4 \phi^4}{1 +
\frac{9(f_0 + f_2 \phi^2 + f_4 \phi^4)}{\beta^2 H^2 \phi^2} }}
\end{eqnarray}
We will consider the dynamics of inflation in two distinct regions:
the first region is when the warp factor is AdS-like
($f(\phi)^{-1}\approx f_4 \phi^4$), and the second is when it is
nearly constant ($f(\phi)^{-1}\approx f_0$).

Starting with AdS region of the throat, we have
\begin{eqnarray}
\label{eq:IRAdSGamma}
\gamma &\approx& \sqrt{1+\frac{1}{9}\frac{\beta^2H^2}{f_4\phi^2}}\\
\dot{\phi} &\approx& \frac{\sqrt{f_4}\phi}{\sqrt{1+\frac{9f_4\phi^2}{\beta^2H^2}}}.
\label{eq:IRAdS}
\end{eqnarray}
For large $\gamma$ one can solve Eq.(\ref{eq:IRAdS}) for the inflaton as a function of time,
\begin{equation}
\phi(t)\approx -\frac{\sqrt{\lambda}}{t}(1-\frac{9}{2\beta^2H^2 t^2}),
\end{equation}
which is the same as previously found in \cite{IRDBI}. As previously
calculated in the same paper, normalization of the density
perturbations requires $\lambda\approx 10^{14}$. This has important
consequences when we consider the size of $\gamma$ in this region.
For the AdS region the motion of the inflaton is relativistic since
$\gamma \approx \frac{\sqrt{\lambda} \beta H}{\phi}$, so for small
$\phi$, $\gamma$ (and thus non-gaussianities) is large.  In fact, we
notice that for generic inflaton masses ($\beta \sim 1$) and the
required value for $\lambda$, we must have $\phi\geq 10^7 H/\gamma$
at 50-60 e-folds back. To be in agreement with non-gaussianity
measurements ($\gamma\approx 20$) and using the upper bound $H\ll
10^{10}$ GeV found in \cite{ChenNonGauss} for IR DBI,
trans-Planckian VEVs can be avoided, as opposed to the UV model.

In the constant region of the throat, we have
\begin{eqnarray}
\gamma &\approx& \sqrt{1+\frac{1}{9}\frac{\beta^2 H^2 \phi^2}{f_0}} =
\sqrt{1+\left(\frac{\phi}{\phi_c}\right)^2} \\
\dot{\phi} &\approx& \sqrt{f_0}\frac{1}{\sqrt{1+9\frac{f_0}{\beta^2 H^2\phi^2}}}
= \sqrt{f_0}\frac{1}{\sqrt{1+\left(\frac{\phi_c}{\phi}\right)^2}}
\label{eq:IRConstGamma}
\end{eqnarray}
where we defined $\phi_c = \frac{3 f_0^{1/2}}{\beta H}$ for
simplicity.  Note that $\frac{\phi_c}{\phi_{tip}}\approx
\frac{\mu}{H\sqrt{\lambda}}$ for the mass gap solution; as noted
above we expect large $\lambda$ to normalize density perturbations
correctly in the AdS region, so we will assume that
$\phi_c\ll\phi_{tip}$. The constant region of the throat can be
further divided up into two regions based on our new scale $\phi_c$:
Region 1 where $0\ll\phi\ll\phi_c$, and Region 2 where
$\phi_c\ll\phi\ll\phi_{tip}$.

     From Eq.(\ref{eq:IRConstGamma}) we see that in Region 1,
$\gamma\approx 1$ and $\phi(t)\sim e^{\beta H t}$ (where
$\phi\rightarrow 0$ as $t\rightarrow -\infty$) so the inflaton is
non-relativistic.  Indeed, explicit calculation of the number of
e-folds and the inflationary parameters indicates that fine tuning
of the inflaton mass $\beta$ is needed to get inflation in this
region,
\begin{eqnarray}
\eta &=& -\frac{\beta}{3} \\
N_e &=& \frac{1}{\eta}\ln(\frac{\phi_{f}}{\phi_{i}}),
\end{eqnarray}
where $\phi_f$ and $\phi_i$ are the starting and ending points of
the inflaton, respectively.  The maximum value of the field in this
region is at the critical value $\phi_c$, while the minimum value is
at a warped string length $\phi_s$.  Taking $f_0^{1/2} = h_{tip}^2
m_s^2$ we have
\begin{equation}
N_e \leq \frac{1}{\eta}\ln(\frac{h_{tip}^2 g_s^{1/2}}{\eta}\frac{m_s}{H})
    \approx \frac{1}{\eta}\ln(\frac{10^{-3}}{\eta}).
\end{equation}
This is the usual slow roll $\eta$ problem (although with stronger
constraints on $\beta$ due to the decreased range for $\phi$) and we
will not investigate this region further.

In Region 2 near the tip of the throat
($\phi_c\ll\phi\ll\phi_{tip}$) the inflaton becomes relativistic
since $\gamma\approx \phi/ \phi_c$.  Using
Eq.(\ref{eq:IRConstGamma}) one can solve for the motion of the
inflaton,
\begin{equation}
\phi(t) = \sqrt{f_0}(t+\frac{9}{2}\frac{1}{\beta^2 H^2 t})
\end{equation}
which is identical to Eq.(2.37) in \cite{ChenNonGauss}.  Here it is
clear that the solution of \cite{ChenNonGauss} is only valid in a
certain region of the constant part of the throat, and can be
combined with the Region 1 solution to obtain a smooth
$\phi\rightarrow 0$ limit.  The number of e-folds and density
perturbations in this region are
\begin{eqnarray}
N_e &=& \int H dt = \frac{H}{\dot{\phi}}\Delta\phi = \frac{H}{\mu}\sqrt{\lambda} \\
\delta_H &=& \frac{H}{M_p}\sqrt{\frac{\gamma}{\epsilon}}
    \approx \left(\frac{H}{\mu}\right)^2\sqrt{\lambda} = \frac{N_e^2}{\sqrt{\lambda}}
\end{eqnarray}
where in the last step we inserted the mass-gap parameters.  It is
possible that in the IR model some e-folds occur in this region,
with the rest of the e-folds occurring in the AdS part of the
throat. One will also generically have problems associated with
large non-gaussianities due to the large value of $\gamma$, which
near $\phi_{tip}$ for the mass gap parameters goes like $\gamma
\approx \frac{H \sqrt{\lambda}}{\mu} $.  As in the AdS case,
however, staying below the upper bound on $H$ allows
non-gaussianities within observational limits.

\section{Discussion}

In this paper we have considered the effects of a cutoff throat on
DBI inflation. We focused on constructions where the brane
spends a significant amount of time in a region of nearly constant warping
near the tip of the throat. To study these types of throats we
assumed that the nearly constant region was larger than a string
length from the tip (Eq.(\ref{eq:TachyonFineTuning})). This may or
may not be more stringent than the requirement that we obtain enough
e-folds (Eq.(\ref{eq:EfoldsTuning})) and the right level of density
perturbations (Eq.(\ref{eq:DensityFineTuning})), since the results
depend on the specific geometry of the warped throat. For a 
weakly warped ($h_{tip}\sim 10^{-2}$)
Klebanov-Strassler throat, we showed that such assumptions are satisfied
\footnote{For more strongly warped KS throats the region of constant warping
is typically dominated by stringy effects.  This can be avoided by
a larger radius for the AdS scale of the throat.}.

In both the UV and IR models of DBI inflation the geometry near
the tip can be important.  In the
former, since the tip is the last region the D-brane experiences
before stringy effects (such as annihilation) become important,
significant inflation in this region can affect inflationary
observables.  In particular, we find that for generically warped
throats 60 e-folds of inflation can happen near the tip, however the
production of large non-gaussian fluctuations seems to be a generic
prediction. From Eq.(\ref{eq:NonGaussThroat}) we see that for a
small enough hierarchy between the Planck and string scale, and a
large enough hierarchy between the inflaton mass and the Planck
scale the non-gaussianities can be sufficiently small.  It is not
clear, however, if this amounts to a fine tuning of the parameters
of the model.

We find that the requirement for enough inflation in the constant
region of the Klebanov-Strassler throat, modeled by the mass gap
approximation, ($R_+/\ell_s \geq 10^4$) is similar to the requirement
that the level of density perturbations for inflation in the AdS
throat yield the correct value ($R_+/\ell_s \sim 10^3-10^4$).  This
implies that it is important to consider inflation at the tip for
UV DBI inflation models.  However, for the mass gap model of the
Klebanov-Strassler throat we find large non-gaussianities, above the
observational limits, for all values of the mass gap parameter $\mu$
consistent with our supergravity analysis.  As discussed above, this
may be avoided by considering other types of throats and
compactifications.

In the IR model of DBI inflation, the geometry near the tip affects
the early time behavior of the inflaton.  In particular, we find
that very near the tip the inflaton is not speed limited (i.e.
$\gamma \approx 1$) and so, without fine tuning of the inflaton
mass, we do not expect inflation there. Further from the tip, but
still in the nearly constant region of the throat, the inflaton
becomes relativistic and the level of non-gaussianities quickly grow
larger. Normalization of density perturbations in both the tip and
AdS regions of the throat requires tuning of the AdS curvature scale
as in the UV model, however agreement with non-gaussianity
observations constrains the Hubble scale of inflation to be smaller
than $10^{13}$ GeV (as is typical for GUT scale inflation).

Throughout this work we have considered only contributions to the
inflaton from the transverse radial mode between the D-branes - it
would be interesting to extend this work to consider the effects of
the angular coordinates on DBI inflation similar to the scenario of
\cite{Giant}. Furthermore, since the inflationary behavior does not
seem to depend on whether we have a D-brane or $\overline{D}$-brane
falling into the throat, one could imagine an $\overline{D}$-brane
attracted to the tip of the throat by the $D3$ charge of the fluxes,
where it collides with a stack of other $\overline{D3}$ already at
the tip. Reheating then occurs in the collision process between the
$\overline{D3}$s. Since we have seen that the D-brane in the UV
model is highly relativistic at the tip of the throat, the
annihilation \cite{Tachyons} and collision process may be
significantly altered along the lines of \cite{Relativistic}, and
may have interesting observational consequences.  In addition, the
formation of cosmic strings from the annihilation of highly
relativistic branes is a relatively unexplored subject, and may
yield different post-annihilation production and properties than the
non-relativistic case \cite{CosmicStrings,CosmicStringReviews}. We
hope to return to these issues in the future.

\section*{Acknowledgments}
It is a pleasure to thank Xingang Chen, Aki Hashimoto, Minxin Huang,
Sarah Shandera, and Henry Tye for discussions and comments. The work
of SK, JM, GS, and BU was supported in part by NSF CAREER Award No.
PHY-0348093, DOE grant DE-FG-02-95ER40896, a Research Innovation
Award and a Cottrell Scholar Award from Research Corporation.

\appendix

\section{The KS Throat}
\setcounter{equation}{0} \label{sec:KSthroat}

The purpose of this section is to show that the mass-gap solution
accurately models the dynamics near the tip, and is a good
qualitative approximation to the full KS throat. For the warped
throat we take the form of the metric used in \cite{Conifolds,Herzog:2001},
\begin{eqnarray}
ds_{10}^2 = \tilde{f}^{-1/2} (\tau) \eta_{\mu \nu} dx^{\mu} dx^{\nu}
+ \tilde{f}^{1/2} (\tau) ds_6^2
\end{eqnarray}
where $\tau$ is the separation between the D-branes in the throat
and the warp factor $\tilde{f}(\tau)$ is defined by
\begin{eqnarray}
\tilde{f}(\tau) &=& 2^{2/3} (g_s M \alpha')^2 \epsilon^{-8/3} I (\tau) \\
I(\tau) &=& \int_{\tau}^{\infty} \frac{x\coth(x)-1}{\sinh^2(x)}
(\sinh(2x)-2x)^{1/3} dx.
\end{eqnarray}
$\epsilon$ is a small real number that is related the deformation of
the warped conifold. If we take $\tau \rightarrow \infty$ then we
recover the AdS approximation, where $\tau$ is redefined in terms of
a radial coordinate $r$. We are interested in keeping the $\tau$
coordinate and observing its behavior near the tip, so the form of
the metric we will use is\footnote{We have suppressed the extra
coordinates of the compact manifold. See \cite{Conifolds} for a
detailed account of the KS metric.}
\begin{eqnarray}
ds_{10}^2 = \tilde{f}^{-1/2} (\tau) \eta_{\mu \nu} dx^{\mu} dx^{\nu} +
\tilde{f}^{1/2} (\tau) \left( \frac{\epsilon^{4/3}}{6 K^2(\tau)} d
\tau^2 + \dots \right)
\end{eqnarray}
where
\begin{eqnarray}
\label{eq:Ktau}
K(\tau) = \frac{(\sinh(2
\tau)-2\tau)^{1/3}}{2^{1/3} \sinh(\tau)}.
\end{eqnarray}

The next step is to calculate the DBI action for the KS throat. To
do this efficiently we will use the definitions:
\begin{eqnarray}
\label{eqn:KSwarp}
h(\tau) &=& \frac{2^{2/3} (g_s M \alpha')^2 I(\tau)\epsilon^{-4/3}}{K^2 (\tau)} \\
\gamma &=& \frac{1}{\sqrt{1-f(\tau) \dot{\tau}^2}}
\end{eqnarray}
The action is then
\begin{eqnarray}
\label{eqn:KSDBI} S_{KS} = -T_{D3} \int d^4 x \sqrt{g}~
\left(\tilde{f}^{-1} (\tau) \left( \sqrt{1-h(\tau) \dot{\tau}^2 } - 1
\right) - V(\tau) \right)
\end{eqnarray}

To consider inflation from the radial coordinate $\tau$ we will define
a canonically normalized scalar field $\varphi$ by expanding the DBI
action for small $\dot{\tau}$,
\begin{eqnarray}
S_{KS} &\approx& \int d^4 x \sqrt{g}\left( \frac{1}{2}T_{D3}
\frac{h(\tau)}{\tilde{f}(\tau)} \dot{\tau}^2 + T_{D3} V(\tau) \right) \\
&=& \int d^4 x \sqrt{g}\left( \frac{1}{2}\dot{\varphi}^2 + V(\varphi)\right)
\end{eqnarray}
where $\tau \approx \frac{K(\tau \rightarrow 0)}{T_{D3}^{1/2}
\epsilon^{2/3}}~\varphi = \left( \frac{2}{3} \right)^{1/3}
\frac{1}{T_{D3}^{1/2} \epsilon^{2/3}}~\varphi$ for small $\tau$.
Rewriting Eq.(\ref{eqn:KSDBI}) in terms of the new scalar field
$\varphi$:
\begin{equation}
S_{KS} = -\int d^4 x \sqrt{g}\left(f(\varphi)^{-1}(\sqrt{1-f(\varphi)\dot{\varphi}^2}-1)
-V(\varphi)\right).
\label{eq:KSDBIScalar}
\end{equation}
This has the same form as the DBI action considered in
Eq.(\ref{eqn:DBIaction}) for the rescaled warp factor $f(\varphi) =
\tilde{f}(\varphi(\tau))/T_{D3}$. In the small $\tau$ expansion we
find $\tilde{f}(\tau)^{-1} \approx 2^{-2/3} (g_s M \alpha')^{-2}
\epsilon^{8/3} (b_0 + b_2 \tau^2 + b_4 \tau^4)$, where $b_0$, $b_2$,
and $b_4$ are constants of ${\cal{O}}(1)$. This warp factor is well
approximated by the mass gap solution after changing to the
canonically normalized field $\varphi$.

\section{Non-Gaussianities}
\label{sec:nongauss}
\setcounter{equation}{0}

In this section we discuss
non-gaussianities in DBI inflation for a generic form of the
warp factor. The results of non-gaussianity for general
single field inflation can be found in
\cite{Chen:2006nt}.  Here we assume the non-gaussianity is large
due to large $\gamma$, and so we can adopt the
method of \cite{Alishahiha:2004eh}.

Using our DBI action (Eq.(\ref{eqn:DBIaction})), our general warp
factor (Eq.(\ref{eq:GeneralF})), a FRW metric for the non-compact
space, and a generic form of the potential $V = V_0 + V_2 \phi^2 +
V_4 \phi^4$, we introduce perturbations to our scalar field
\begin{eqnarray}
\phi \rightarrow \phi(t) + \alpha(x,t).
\end{eqnarray}
Non-gaussianities come from the third order interactions in our
Lagrangian due to the perturbation $\alpha(x,t)$,
\begin{eqnarray}
\label{eqn:genL3} {\cal{L}}_3 &=& -a(t)^3 \left[ \frac{\gamma^5
\dot{\phi}}{2 (f_0 + f_2 \phi^2 + f_4 \phi^4) } \dot{\alpha}^3 -
\frac{ \gamma^3 \dot{\phi} }{2 a^2 (f_0 + f_2 \phi^2 + f_4 \phi^4)}
\dot{\alpha} (\nabla \alpha)^2 \right. \\ \nonumber && \left. +
\frac{\gamma^3 \phi (f_2 + 2 f_4 \phi^2) \dot{\phi}^2
}{2 a^2 (f_0 + f_2 \phi^2 + f_4 \phi^4)^2} \alpha (\nabla \alpha)^2 \right. \\
\nonumber && \left. + \frac{\gamma^3 \dot{\phi}^3 }{2 (f_0 + f_2
\phi^2 + f_4 \phi^4 )^4 } \left( -f_0^2 f_2 + 19 f_2 f_4^2 \phi^8 +
10 f_4^3 \phi^{10} \right. \right. \\ \nonumber && \left. \left. +
\phi^2 (2 f_0 (f_2^2 - 3 f_0 f_4) + 3 f_2^2 \gamma^2 \dot{\phi}^2 )
+ 4 f_4 \phi^6 (3 f_2^2 + f_0 f_4 + 3 f_4 \gamma^2 \dot{\phi}^2)
\right. \right. \\ \nonumber && \left. \left. + f_2 \phi^4 (3 f_2^2
+ 2 f_0 f_4 + 12 f_4 \gamma^2 \dot{\phi}^2) \right) \dot{\alpha}
\alpha^2 - \frac{3
\gamma^5 \phi (f_2 + 2 f_4 \phi^2) \dot{\phi}^2}{2(f_0 + f_2 \phi^2 + f_4 \phi^4)^2} \dot{\alpha}^2 \alpha \right. \\
\nonumber && \left. + \frac{\phi}{2 \gamma} \left( 8 (f_4 (-1 +
\gamma) - V_4 \gamma) - \frac{4 f_4 \gamma^2 \dot{\phi}^2 }{f_0 +
f_2 \phi^2 + f_4 \phi^4} \right. \right. \\ \nonumber && \left.
\left. + \frac{\gamma^4 \dot{\phi}^4 }{(f_0 + f_2 \phi^2 + f_4
\phi^4)^4} (f_2 + 2 f_4 \phi^2)(f_0 f_2 - \phi^2 (f_2^2 - 6 f_0 f_4
+ f_4 \phi^2 (f_2 + 2 f_4 \phi^2))) \right. \right. \\ \nonumber &&
\left. \left. - \frac{\gamma^6 \phi^2 \dot{\phi}^6 (f_2 + 2 f_4
\phi^2)^3 }{(f_0 + f_2 \phi^2 + f_4 \phi^4)^5} \right) \alpha^3
\right]
\end{eqnarray}

The behavior of the non-gaussian fluctuations will be dominated by
the small $\phi$ (alternatively, late time) behavior of the
perturbations near the tip of the throat. Plugging our value for
$\dot{\phi}$ from Eq.(\ref{phidot}), we can evaluate the leading
$\phi$ behavior of ${\cal{L}}_3$. Of the terms in
Eq.(\ref{eqn:genL3}), the $\dot{\alpha}^3$ and $\dot{\alpha}
\nabla{\alpha}^2$ are dominant for small $\phi$, and produce the
same results as in \cite{Alishahiha:2004eh, ChenNonGauss}. The
results for these terms hold for all choices of warp factor.

Naively the first term from the $\dot{\alpha} \alpha^2$
contribution ($\propto -f_0^2 f_2$) can also possibly contribute to
the non-gaussianities. Using the procedure outlined in
\cite{Alishahiha:2004eh}
and evaluating this term for the mass gap
solution, we find that it produces non-gaussianities of
${\cal{O}}(\gamma)$, which is subleading in the limit
of large $\gamma$.  Since the procedure in \cite{Alishahiha:2004eh}
only produces the leading non-gaussianities, ${\cal{O}}(\gamma)$
contributions will be dropped.


\begin{thebibliography}{99}



\bibitem{DvaliTye}
    G. Dvali, S.H.H. Tye,
    Phys. Lett. B. {\bf 450}, 72 (1999), [arXiv:hep-th/9812483].

\bibitem{DD}
    C.P. Burgess, M. Majumdar, D. Nolte, F. Quevedo, G. Rajesh, R.J. Zhang,
    JHEP {\bf 0107},
    047 (2001), [arXiv:hep-th/0105204];
    G.~R.~Dvali, Q.~Shafi and S.~Solganik,
  [arXiv:hep-th/0105203];
 G.~Shiu and S.~H.~H.~Tye,
  Phys.\ Lett.\ B {\bf 516}, 421 (2001)
  [arXiv:hep-th/0106274];
    C.P. Burgess, P. Martineau, F. Quevedo, G. Rajesh and R. J. Zhang,
    JHEP {\bf 0203}, 052 (2002) [arXiv:hep-th/0111025];
    H. Firouzjahi and S. H. H. Tye,
    Phys. Lett. B {\bf 584}, 147 (2004) [arXiv:hep-th/0312020];
    C. P. Burgess, J. M. Cline, H. Stoica and F. Quevedo,
    JHEP {\bf 0409}, 033 (2004) [arXiv:hep-th/0403119];
    N. Iizuka, S.P. Trivedi,
    Phys.Rev. D{\bf 70} (2004) 043519 [arXiv:hep-th/0403203];
    A. Buchel, A. Ghodsi,
    Phys.Rev. D{\bf 70} (2004) 126008 [arXiv:hep-th/0404151];
    A. Buchel,
    [arXiv:hep-th/0601013].


\bibitem{KKLMMT}
    S. Kachru, R. Kallosh, A. Linde, J. Maldacena, L. McAllister, S. Trivedi,
    JCAP {\bf 0310} (2003) 013
    [arXiv:hep-th/0308055].

\bibitem{AngleInflation}
    J. Garcia-Bellido, R. Rabadan, R. Zamora,
    JHEP {\bf 0201}, 036 (2002), [arXiv:hep-th/0112147];
    R. Blumenhagen, B. Kors, D. Lust and T. Ott,
    Nucl. Phys. B {\bf 641}, 235 (2002) [arXiv:hep-th/0202124];
    N. Jones, H. Stoica, S.H.H. Tye,
    JHEP {\bf 0207}, 051 (2002), [arXiv:hep-th/0203163];
     M.~Gomez-Reino and I.~Zavala,
  JHEP {\bf 0209}, 020 (2002)
  [arXiv:hep-th/0207278].
    S. Shandera, B. Shaler, H. Stoica, S.H.H. Tye,
    JCAP {\bf 0402}, 013 (2004), [arXiv:hep-th/0311207].

\bibitem{StringInflation}
    F. Quevedo,
    Class. Quant. Grav 19, 5721-5779C (2002) [arXiv:hep-th/0210292];
    K. Dasgupta, C. Herdeiro, S. Hirano and R. Kallosh,
    Phys. Rev. D {\bf 65}, 126002 (2002) [arXiv:hep-th/0203019];
    G.~Shiu and I.~Wasserman,
    Phys.\ Lett.\ B {\bf 541}, 6 (2002)
    [arXiv:hep-th/0205003];
    G.~Shiu, S.~H.~H.~Tye and I.~Wasserman,
    Phys.\ Rev.\ D {\bf 67}, 083517 (2003)
    [arXiv:hep-th/0207119];
    J. J. Blanco-Pillado et al.,
    JHEP {\bf 0411}, 063 (2004) [arXiv:hep-th/0406230];
    J. J. Blanco-Pillado et al.,
    [arXiv:hep-th/0603129];
    H. Firouzjahi and S. H. H. Tye,
    JCAP {\bf 0503}, 009 (2005) [arXiv:hep-th/0501099];
    K. Becker, M. Becker and A. Krause,
    Nucl. Phys. B {\bf 715}, 349 (2005) [arXiv:hep-th/0501130];
    A. Westphal,
    JCAP 0511 (2005) 003 [arXiv:hep-th/0507079];
    E. Papantonopoulos, I. Pappa, V. Zamarias,
    [arXiv:hep-th/0601152].


\bibitem{Giant}
  O.~DeWolfe, S.~Kachru and H.~L.~Verlinde,
  JHEP {\bf 0405}, 017 (2004)
  [arXiv:hep-th/0403123].




\bibitem{Reheating}
G.~Shiu, S.~H.~H.~Tye and I.~Wasserman,
  Phys.\ Rev.\ D {\bf 67}, 083517 (2003)
  [arXiv:hep-th/0207119];
    N. Barnaby, C. P. Burgess and J. M. Cline,
    JCAP 0504, 007 (2005),
    [arXiv:hep-th/0412040];
    D.~Chialva, G.~Shiu and B.~Underwood,
  JHEP {\bf 0601}, 014 (2006)
  [arXiv:hep-th/0508229];
    A. R. Frey, A. Mazumdar and R. Myers,
    Phys. Rev. D{\bf 73}, 026003 (2006),
    [arXiv:hep-th/0508139];
  X. Chen and S. H. H. Tye,
    [arXiv:hep-th/0602136];
    P. Langfelder,
    [arXiv:hep-th/0602296].


\bibitem{Kofman:2005yz}
  L.~Kofman and P.~Yi,
  Phys.\ Rev.\ D {\bf 72}, 106001 (2005)
  [arXiv:hep-th/0507257].

\bibitem{CosmicStrings}
S. Sarangi and S. H. H. Tye,
Phys. Lett. B {\bf 536}, 185 (2002) [arXiv:hep-th/0204074];
G.~Shiu,
[arXiv:hep-th/0210313];
N. T. Jones, H. Stoica and S. H. H. Tye,
Phys. Lett. B {\bf 563}, 6 (2003) [arXiv:hep-th/0303269];
 L. Pogosian, S. H. H. Tye, I. Wasserman and M. Wyman,
 Phys. Rev. D {\bf 68}, 023506 (2003) [arXiv:hep-th/0304188];
 T. Damour and A. Vilenkin,
 Phys. Rev. D {\bf 71}, 063510 (2005) [arXiv:hep-th/0410222];
  N. Barnaby, A. Berndsen, J. M. Cline and H. Stoica,
  JHEP {\bf 0506}, 075 (2005) [arXiv:hep-th/0412095];
  G. Dvali, R. Kallosh and A. Van Proeyen,
  JHEP {\bf 0401} (2004) 035 [hep-th/0312005];
  G. Dvali and A. Vilenkin,
  JCAP {\bf 0403} (2004) 010 [hep-th/0312007];
  K. Dasgupta, J. P. Hsu, R. Kallosh, A. Linde and M. Zagermann,
  [hep-th/0405247];
  B. Chen, M. Li, J. She,
  JHEP {\bf 0506} (2005) 009,
  [arXiv:hep-th/0504040];
  E.~J.~Copeland, R.~C.~Myers and J.~Polchinski,
  JHEP {\bf 0406}, 013 (2004)
  [arXiv:hep-th/0312067].


\bibitem{Silverstein:2003hf}
  E.~Silverstein and D.~Tong,
  Phys.\ Rev.\ D {\bf 70}, 103505 (2004)
  [arXiv:hep-th/0310221].

\bibitem{Alishahiha:2004eh}
  M.~Alishahiha, E.~Silverstein and D.~Tong,
  Phys.\ Rev.\ D {\bf 70}, 123505 (2004)
  [arXiv:hep-th/0404084].

\bibitem{IRDBI} X. Chen,
    Phys.Rev. D{\bf 71} (2005) 063506
    [arXiv:hep-th/0408084];
    JHEP {\bf 0508} (2005) 045
    [arXiv:hep-th/0501184].


\bibitem{ChenNonGauss}
    X. G. Chen,
    [arXiv:astro-ph/0507053].




\bibitem{Herman}
H. Verlinde, Nucl. Phys. {\bf B580} (2000) 264, [arXiv:hep-th/9906182].

\bibitem{Gukov:1999ya}
  S.~Gukov, C.~Vafa and E.~Witten,
  ``CFT's from Calabi-Yau four-folds,''
  Nucl.\ Phys.\ B {\bf 584}, 69 (2000)
  [Erratum-ibid.\ B {\bf 608}, 477 (2001)]
  [arXiv:hep-th/9906070].

\bibitem{Dasgupta:1999ss}
  K.~Dasgupta, G.~Rajesh and S.~Sethi,
  ``M theory, orientifolds and G-flux,''
  JHEP {\bf 9908}, 023 (1999)
  [arXiv:hep-th/9908088].

\bibitem{Greene:2000gh}
  B.~R.~Greene, K.~Schalm and G.~Shiu,
  ``Warped compactifications in M and F theory,''
  Nucl.\ Phys.\ B {\bf 584}, 480 (2000)
  [arXiv:hep-th/0004103].


\bibitem{GKP} S.~B.~Giddings, S.~Kachru and J.~Polchinski,
  Phys.\ Rev.\ D {\bf 66}, 106006 (2002)
  [arXiv:hep-th/0105097].

\bibitem{KKLT}
    S. Kachru, R. Kallosh, A. Linde, S.P. Trivedi,
    Phys. Rev. D, {\bf 68}, 046005 (2003), [arXiv:hep-th/0301240].

\bibitem{Klebanov:2000hb}
  I.~R.~Klebanov and M.~J.~Strassler,
  JHEP {\bf 0008}, 052 (2000)
  [arXiv:hep-th/0007191].

\bibitem{Klebanov:2000nc}
  I.~R.~Klebanov and A.~A.~Tseytlin,
  Nucl.\ Phys.\ B {\bf 578}, 123 (2000)
  [arXiv:hep-th/0002159].

\bibitem{Conifolds}
    P. Candelas and X. C. de la Ossa,
    Nucl. Phys. B {\bf 342}, 246 (1990);
    R. Minasian and D. Tsimpis,
    Nucl. Phys. B {\bf 572}, 499 (2000),
    [arXiv:hep-th/9911042];
    K. Ohta and T. Yokono,
    JHEP {\bf 0002}, 023 (2000),
    [arXiv:hep-th/9912266];
    H.~Firouzjahi and S.~H.~H.~Tye,
  JHEP {\bf 0601}, 136 (2006)
  [arXiv:hep-th/0512076].

\bibitem{Herzog:2001}
    C. P. Herzog, I. R. Klebanov and P. Ouyang,
    [arXiv:hep-th/0108101];

\bibitem{Shandera:2006ax}
  S.~E.~Shandera and S.~H.~H.~Tye,
  [arXiv:hep-th/0601099].





\bibitem{WMAP}
    D.N. Spergel et. al,
    [arXiv:astro-ph/0603449].


\bibitem{HamJac}
  A.~G.~Muslimov,
  Class.\ Quant.\ Grav.\  {\bf 7}, 231 (1990);
  D.~S.~Salopek and J.~R.~Bond,
  Phys.\ Rev.\ D {\bf 42}, 3936 (1990);
  J.~E.~Lidsey,
  Phys.\ Lett.\ B {\bf 273}, 42 (1991).

\bibitem{Shandera:2004zy}
    R. D'Auria, S. Ferrara and M. Trigiante,
    [arXiv:hep-th/0407138];
    L. Andrianopoli, R. D'Auria, S. Ferrara and M. A. Lledo,
    JHEP {\bf 0303}, 044 (2003)
    [arXiv:hep-th/0302174];
    P. K. Tripathy and S. P. Trivedi,
    JHEP {\bf 0303}, 028 (2003)
    [arXiv:hep-th/0301139];
    C. Angelantonj, R. D'Auria, S. Ferrara and M. Trigiante,
    Phys. Lett. B {\bf 583}, 331 (2004)
    [arXiv:hep-th/0312019];
    A. R. Frey and J. Polchinski,
    Phys. Rev. D {\bf 65}, 126009 (2002)
    [arXiv:hep-th/0201029];
    S. Kachru, M. B. Schulz and S. Trivedi,
    JHEP {\bf 0310}, 007 (2003)
    [arXiv:hep-th/0201028].
  S.~E.~Shandera,
  JCAP {\bf 0504}, 011 (2005)
  [arXiv:hep-th/0412077].

\bibitem{WarpedEFT}
O. DeWolfe and S.B. Giddings, Phys. Rev. {\bf D67} (2003) 066008,
[arXiv:hep-th/0208123]; S.B. Giddings and A. Maharana, [arXiv:hep-th/0507158];
A. Frey and A. Maharana, [arXiv:hep-th/0603233].


\bibitem{Kodama}
    H. Kodama and K. Uzawa,
    JHEP {\bf 0507}, 061 (2005)
    [arXiv:hep-th/0504193].

\bibitem{FluxAnnihilation}
    S. Kachru, J. Pearson and H. L. Verlinde,
    JHEP {\bf 0206}, 021 (2002)
    [arXiv:hep-th/0112197].

\bibitem{RS}
    L.~Randall and R.~Sundrum,
  Phys.\ Rev.\ Lett.\  {\bf 83}, 3370 (1999)
  [arXiv:hep-ph/9905221].

\bibitem{Tachyons}
    P. Yi, Nucl. Phys. B {\bf 550}, 214 (1999) [arXiv:hep-th/9901159];
    O. Bergman, K. Hori and P. Yi, Nucl. Phys. B {\bf 580}, 289 (2000) [arXiv:hep-th/0002223];
    G. W. Gibbons, K. Hori and P. Yi, Nucl. Phys. B {\bf 596}, 136 (2001) [arXiv:hep-th/0009061];
    A. Sen,
    J. Math. Phys. {\bf 42}, 2844 (2001) [arXiv:hep-th/0010240];
    G. Gibbons, K. Hashimoto and P. Yi, JHEP {\bf 0209}, 061 (2002) [arXiv:hep-th/0209034];
    B. Chen, M. Li and F. L. Lin, JHEP {\bf 0211}, 050 (2002) [arXiv:hep-th/0209222];
    P. Mukhopadhyay and A. Sen, JHEP {\bf 0211}, 047 (2002) [arXiv:hep-th/0208142];
    S. J. Rey and S. Sugimoto, Phys. Rev. D {\bf 67}, 086008 (2003) [arXiv:hep-th/0301049];
    A. Sen, JHEP {\bf 0204}, 048 (2002) [arXiv:hep-th/0203211]; JHEP {\bf 0207}, 065 (2002) [arXiv:hep-th/0203265];
    Mod. 
Phys. Lett. A {\bf 17}, 1797 (2002) [arXiv:hep-th/0204143]; Phys. Rev. Lett. {\bf 91}, 181601 (2003)
[arXiv:hep-th/0306137]; 
[arXiv:hep-th/0410103];
    N. Lambert, J. Liu, J. Maldacena,
    [arXiv:hep-th/0303139];
    M. Gutperle and P. Yi, JHEP {\bf 0501}, 015 (2005)
[arXiv:hep-th/0409050]; J. Shelton,
JHEP {\bf 0501} (2005) 037 [arXiv:hep-th/0411040];
    N.T. Jones, L. Leblond, S.H.H. Tye,
    JHEP {\bf 0310} (2003) 002,
    [arXiv:hep-th/0307086].
    X. Chen,
    Phys.Rev. D{\bf 70} (2004) 086001
    [arXiv:hep-th/0311179].


\bibitem{Relativistic}
    L. McAllister and I. Mitra,
    JHEP {\bf 0502}, 019 (2005) [arXiv:hep-th/0408085].


  \bibitem{CosmicStringReviews}
For some recent reviews, see, e.g.,  J.~Polchinski,
  [arXiv:hep-th/0412244];
 T.~W.~B.~Kibble,
  [arXiv:astro-ph/0410073];
  A.~Vilenkin,
  [arXiv:hep-th/0508135].


\bibitem{Chen:2006nt}
  X.~Chen, M.~x.~Huang, S.~Kachru and G.~Shiu,
  arXiv:hep-th/0605045.
\end{thebibliography}
\end{document}